\documentclass[sigconf]{acmart}
\AtBeginDocument{%
  }
\usepackage{subcaption} 
\copyrightyear{2026}
\acmYear{2026}
\setcopyright{cc}
\setcctype{by}
\acmConference[CHI EA '26]{Extended Abstracts of the 2026 CHI Conference on Human Factors in Computing Systems}{April 13--17, 2026}{Barcelona, Spain}
\acmBooktitle{Extended Abstracts of the 2026 CHI Conference on Human Factors in Computing Systems (CHI EA '26), April 13--17, 2026, Barcelona, Spain}
\acmDOI{10.1145/3772363.3798940}
\acmISBN{979-8-4007-2281-3/2026/04}

\begin{document}

\title{Cognitive Prosthetic: An AI-Enabled Multimodal System for Episodic Recall in Knowledge Work}

\author{Lawrence Obiuwevwi}
\authornote{Corresponding author.}
\email{lobiu001@odu.edu}
\affiliation{%
  \institution{Old Dominion University}
  \city{Norfolk}
  \state{Virginia}
  \country{United States}
}

\author{Krzysztof J. Rechowicz}
\email{krechowi@odu.edu}
\affiliation{%
  \institution{Old Dominion University}
  \city{Norfolk}
  \state{Virginia}
  \country{United States}
}

\author{Vikas Ashok}
\email{vganjiqu@cs.odu.edu}
\affiliation{%
  \institution{Old Dominion University}
  \city{Norfolk}
  \state{Virginia}
  \country{United States}
}

\author{Sachin Shetty}
\email{sshetty@odu.edu}
\affiliation{%
  \institution{Old Dominion University}
  \city{Norfolk}
  \state{Virginia}
  \country{United States}
}

\author{Sampath Jayarathna}
\authornote{Corresponding author.}
\email{sampath@cs.odu.edu}
\affiliation{%
  \institution{Old Dominion University}
  \city{Norfolk}
  \state{Virginia}
  \country{United States}
}

\renewcommand{\shortauthors}{Lawrence et al.}


\begin{abstract}
Modern knowledge workplaces increasingly strain human episodic memory as individuals navigate fragmented attention, overlapping meetings, and multimodal information streams. Existing workplace tools provide partial support through note-taking or analytics but rarely integrate cognitive, physiological, and attentional context into retrievable memory representations. This paper presents the \textbf{Cognitive Prosthetic Multimodal System (CPMS)}—an AI-enabled proof-of-concept designed to support episodic recall in knowledge work through structured episodic capture and natural language retrieval. CPMS synchronizes speech transcripts, physiological signals, and gaze behavior into temporally aligned, JSON-based episodic records processed locally for privacy. Beyond data logging, the system includes a web-based retrieval interface that allows users to query past workplace experiences using natural language, referencing semantic content, time, attentional focus, or physiological state. We present CPMS as a functional proof-of-concept demonstrating the technical feasibility of transforming heterogeneous sensor data into queryable episodic memories. The system is designed to be modular, supporting operation with partial sensor configurations, and incorporates privacy safeguards for workplace deployment. This work contributes an end-to-end, privacy-aware architecture for AI-enabled memory augmentation in workplace settings.
\end{abstract}





\begin{teaserfigure}
    \centering
\includegraphics[width=\textwidth,height=0.45\textheight,keepaspectratio]{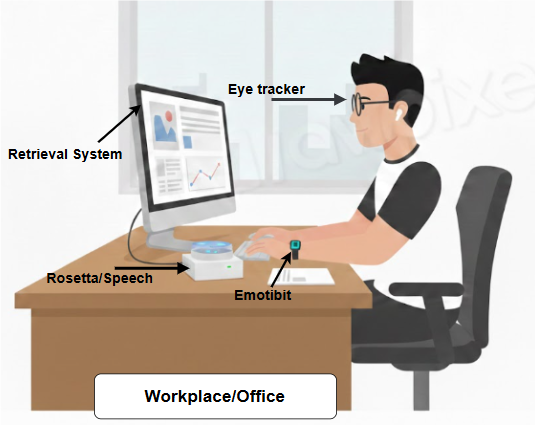}
    \Description{Teaser figure showing the system at a glance.}
    \caption{Cognitive Prosthetic Multimodal System (CPMS).}
    \label{fig:teaser}
\end{teaserfigure}

\maketitle

\begin{figure*}[t!]
    \centering
    \includegraphics[width=\linewidth,height=8.000cm]{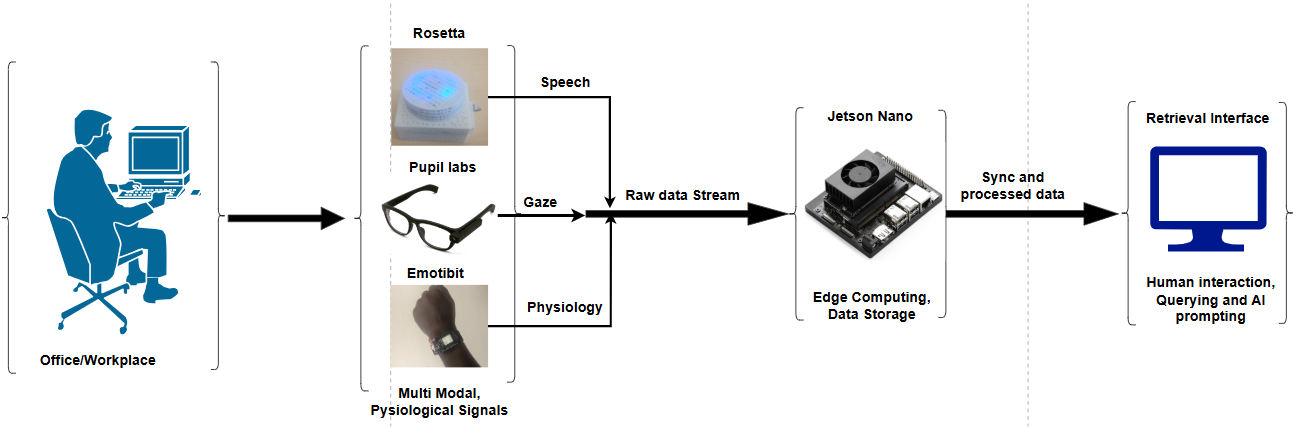}
    \Description{Block diagram of the CPMS pipeline from sensing to episodic encoding, storage, and natural language retrieval.}
    \caption{CPMS High-level architecture illustrating multimodal sensing, episodic encoding into structured memory records, and AI-enabled natural language retrieval for workplace reflection.}
    \label{fig:architectures}
\end{figure*}

\section{Introduction}

Contemporary knowledge work increasingly unfolds across fragmented meetings, digital tools, and parallel tasks, placing sustained pressure on human episodic memory. Workers are frequently required to recall what was said, where attention was directed, and how decisions unfolded, often under conditions of stress or cognitive overload. Human memory, however, is not well suited for reconstructing such distributed experiences, leading to recall gaps, context loss, and inefficiencies in task resumption \cite{yang2021, epstein2020}.

Early digital memory aids explored lifelogging as a means of supporting recall, demonstrating that passively captured visual cues could trigger autobiographical memory \cite{zacks2020, baldwin2020}. Subsequent personal informatics systems expanded this approach to multimodal data streams, including activity and health signals \cite{cho2022, heersmink2022}. While these efforts showed promise, they largely treated modalities in isolation and rarely supported integrated, context-aware retrieval suited for professional environments \cite{yen2021}.

In parallel, workplace sensing research has matured across speech analysis \cite{jayawardena2020pilot}, eye tracking \cite{thennakoon2025devices}, and physiological monitoring \cite{de2019survey}. Speech transcription and diarization enable reconstruction of meetings and conversations \cite{song2021}, eye tracking \cite{thennakoon2025beyond} provides insight into attention and problem-solving strategies \cite{agustianto2022, turkoglu2024, pourhosein2025}, and physiological signals such as heart rate and galvanic skin response reveal stress and cognitive load \cite{midha2020, wang2021}. However, these modalities are typically analyzed independently, limiting their usefulness for holistic memory reconstruction.

Recent advances in AI-driven augmentation, particularly large language models (LLMs), suggest new opportunities for integrating heterogeneous data into accessible representations. AI systems have been shown to reduce mental workload and improve task support in professional contexts \cite{colabianchi2024, duchevet2022, brynjolfsson2023}. Yet, most such systems are reactive, offering momentary assistance rather than constructing enduring, queryable records of experience.

To address this gap, we present the \textit{Cognitive Prosthetic Multimodal System (CPMS)}—an AI-enabled proof-of-concept designed to support episodic recall in knowledge work through structured episodic capture and natural language retrieval. CPMS synchronizes speech, physiological signals, and gaze behavior into temporally aligned, JSON-based memory logs. These logs form a unified episodic archive that can be accessed through a web-based interface supporting natural language queries across time, content, attention, and physiological state.

This paper presents CPMS as a functional proof-of-concept for AI-enabled episodic memory augmentation, demonstrating the technical feasibility of multimodal capture, temporal synchronization, and natural language retrieval rather than evaluating cognitive outcomes. The system is designed with a modular architecture that supports operation under partial sensor availability, and incorporates privacy safeguards appropriate for workplace deployment. The remainder of the paper describes related work, system design, implementation details, ethical considerations, and implications for future cognitive prosthetic systems.
\section{Related Work}

Digital memory augmentation has long explored how technology can support human episodic recall. Early lifelogging systems such as SenseCam and Memoto demonstrated that passively captured visual cues could meaningfully trigger autobiographical memory, particularly for individuals with memory impairments \cite{yang2021, zacks2020, baldwin2020}. Personal informatics research later expanded this paradigm to include activity, location, and health data, enabling reflective analysis of daily behavior \cite{jayarathna2017analysis, epstein2020, cho2022}. However, these systems typically focused on lifestyle tracking and personal reflection rather than the cognitive demands of professional work, and they rarely integrated heterogeneous modalities into unified representations \cite{jayarathna2016rationale, yen2021, heersmink2022}.

More recent lifelogging efforts emphasize semantic interpretation and contextual grounding. Approaches such as semantic lifelogging frameworks and large-scale egocentric datasets aim to move beyond raw capture toward meaning-centered representations \cite{ali2022, grauman2022}. Gaze-driven \cite{abeysinghe2025framework} and attention-aware methods \cite{jayawardena2025real} further link visual focus to episodic recall and task understanding \cite{gonzalez2024, Kumushini2025}. While these advances highlight the value of multimodal context, they often prioritize dataset scale or algorithmic benchmarking over deployable, privacy-aware systems for everyday workplace use \cite{yen2021}.

In parallel, research on workplace sensing has produced robust pipelines for capturing speech, physiological signals, and attentional behavior. Speech transcription and diarization support reconstruction of meetings and collaborative interactions \cite{song2021}. Eye tracking has been widely applied to study attention, problem-solving strategies, and communication effectiveness in professional settings \cite{turkoglu2024, pourhosein2025}. Physiological sensing \cite{obiuwevwi2025toward}, including heart rate and galvanic skin response, provides insight into stress, workload, and affective states during work tasks \cite{izumi2021, ding2022, obiuwevwi2025hypoglycemia}. Although these modalities are individually well studied, they are most often analyzed in isolation, limiting their ability to support holistic episodic recall.

Cognitive psychology offers additional insight through theories of event segmentation, which show that human memory organizes continuous experience into discrete, meaningful episodes \cite{zacks2020, baldwin2020}. Computational systems inspired by this principle increasingly structure multimodal data into event-based representations, using graphs or hybrid schemas to support contextual retrieval \cite{ribeiro2023, Hodges2006}.

 Recent work further demonstrates how event boundaries influence emotional memory and reduce interference between experiences \cite{laing2025, li2025}, motivating systems that align computational segmentation with human memory structure.

At the intersection of these threads, augmented cognition and assistive prosthetics research explores how multimodal and affect-aware systems can extend human cognitive capacity. Neuroprosthetic and affective computing systems highlight the importance of integrating behavioral and physiological signals while preserving interpretability and user trust \cite{belkacem2020, calvo2021affect}. Advances in language models and multimodal learning further enable semantic fusion and natural language interaction with complex data streams \cite{zhou2024}. Yet, few systems combine these advances into unified, privacy-conscious architectures that explicitly support retrospective workplace memory rather than real-time task assistance.

The Cognitive Prosthetic Multimodal System (CPMS) builds on these foundations by unifying speech, gaze, and physiological signals into structured, episodic memory logs designed for interactive retrieval. Unlike exhaustive lifelogging or single-modality sensing systems, CPMS emphasizes selective capture, temporal alignment, and interpretability, positioning it as a pragmatic step toward deployable cognitive prosthetics for professional environments.

\section{Methodology}

The CPMS is implemented as an end-to-end multimodal episodic memory system that combines synchronized sensing, structured archival storage, and AI-enabled natural language retrieval for workplace environments. Unlike real-time monitoring or intervention systems \cite{izumi2021, ding2022}, CPMS adopts a retrospective, memory-oriented design in which multimodal signals are captured, aligned, and stored as queryable episodic records. Building on lifelogging research showing that passive capture can cue episodic recall \cite{yang2021, zacks2020, baldwin2020}, CPMS extends prior work by enabling direct natural language access to multimodal workplace experiences through a dedicated memory retrieval web application.

The system integrates three complementary sensing modalities—speech, physiological signals, and gaze behavior—captured using wearable and ambient devices (Figure~\ref{fig:architectures}).
Speech is continuously recorded using the Storymodelers Rosetta Stone device and transcribed via automatic speech recognition systems such as Whisper or Google Speech-to-Text \cite{song2021, tilkar2025a}. Physiological signals, including heart rate and galvanic skin response (GSR), are captured using the EmotiBit wearable, providing validated indicators of cognitive load and affective arousal \cite{izumi2021, ding2022, obiuwevwi2025hypoglycemia}. Gaze behavior is recorded using the Pupil Labs Core eye tracker, with fixation, blink, and saccade events aggregated into one-second summaries reflecting attentional focus \cite{turkoglu2024, pourhosein2025}. Because these modalities operate at different sampling rates \cite{jayawardana2022streaminghub}, all streams are temporally aligned to a shared one-second reference grid, resampled, and aggregated into discrete episodic units that encode semantic content, physiological state, and attentional behavior within each time slice, operationalizing theories of event segmentation \cite{zacks2020, baldwin2020}.

\subsection{Modular Design of CPMS.} 
CPMS operates under partial sensor availability by processing each modality independently before temporal alignment. Episodic records are generated even when certain sensors are absent or noisy, enabling flexible deployment across workplace settings. For example, if physiological data are unavailable, synchronized speech and gaze records are still archived.

All processing and storage occur locally on an NVIDIA Jetson Nano to support privacy-by-design and reduce cloud reliance \cite{liu2023, badidi2023}. The system follows a four-layer architecture: ingestion, synchronization and processing, archival storage, and retrieval (see Figure~\ref{fig:architectures}). Records are serialized into structured JSON/JSONL formats containing transcript segments, normalized physiological signals, and gaze indicators \cite{dobbins2017detecting, schmidt2021}.

\subsection{Retrieval Mechanism.} 
Archived records are accessed through an LLM-powered web interface. Queries are resolved using hybrid retrieval: filtering by temporal and physiological metadata \cite{jayawardana2021metadata,jayawardana2019dfs} followed by semantic matching over transcript content. Results return transcript excerpts with timestamps and associated physiological and gaze context.

For example, a query about decisions made during high stress filters by time and elevated physiological thresholds before retrieving relevant transcript segments. Limitations include ASR noise, gaze ambiguity, and motion artifacts. CPMS thus functions as an AI-enabled cognitive prosthetic supporting reflective episodic recall \cite{gupta2022, Marusich2025}.

\section{Results}

\begin{figure}[H]
    \centering
    \includegraphics[width=\linewidth,keepaspectratio]{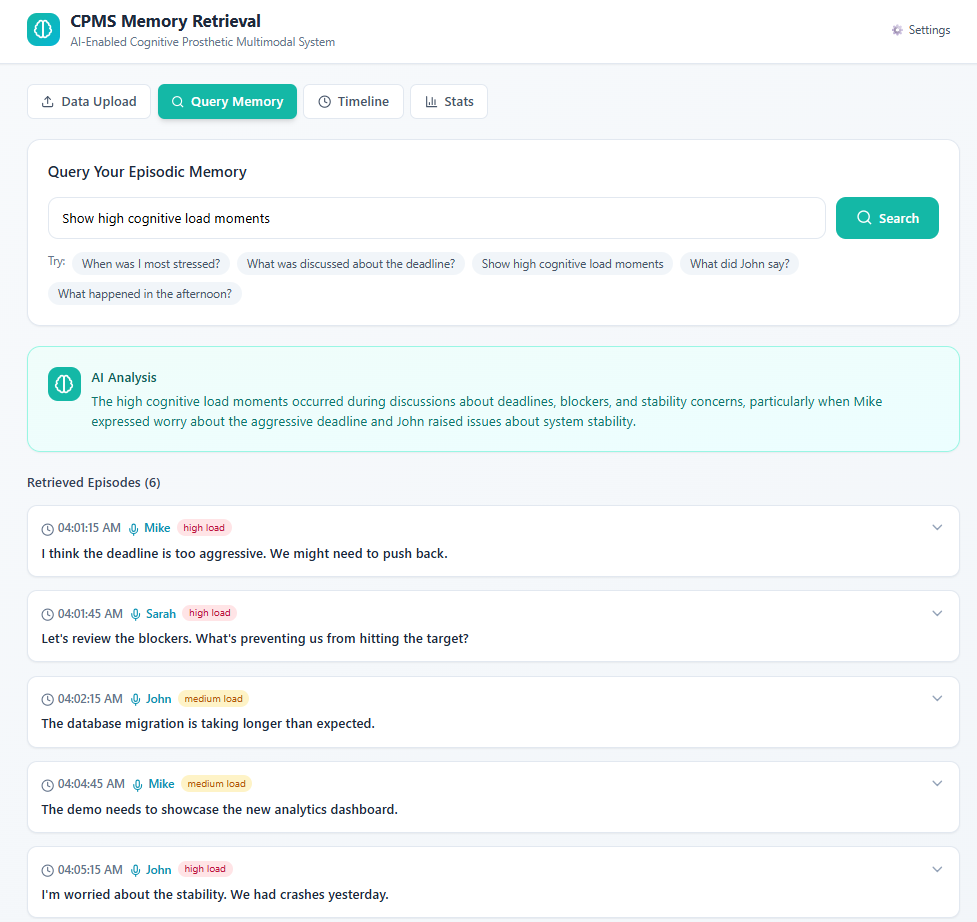}
    \Description{Screenshot of the natural language query interface with a text box and returned results.}
    \caption{Natural language query interface for episodic retrieval.}
    \label{fig:cpms-interface}
\end{figure}

The CPMS system successfully captured, synchronized, and archived multimodal workplace signals into a unified, machine-readable episodic memory representation. Speech transcripts, physiological signals, and gaze events were aligned to a shared one-second temporal grid, producing coherent multimodal ``memory slices'' that jointly encode semantic content, affective state, and attentional context. Figure~\ref{fig:cpms-interface} presents representative outputs from the speech and physiological modalities, illustrating data output required for structured episodic logging and the queryable retrieval webapp.

All synchronized episodes were serialized into structured JSON/JSONL records representing one-second multimodal snapshots. Each record integrates spoken content with normalized physiological measures (e.g., heart rate and galvanic skin response) and gaze indicators, enabling cross-modal association within a shared timeline. This transformation converts continuous, high-bandwidth sensor streams into lightweight, interpretable representations suitable for semantic indexing and AI-assisted retrieval, while preserving essential temporal and contextual fidelity \cite{liu2023, xu2023multimodal}.

Beyond data capture and archival, CPMS demonstrated functional episodic retrieval through its web-based memory interface. In proof-of-concept testing, natural language queries referencing semantic content, temporal context, attentional focus, or physiological state (e.g., ``What was discussed during moments of elevated stress?'') successfully retrieved corresponding episodic segments from the archive. The system filtered records using temporal and physiological metadata before applying semantic matching over transcript content, returning relevant excerpts alongside timestamps and associated contextual signals. These results indicate that multimodal episodic records can be meaningfully indexed and accessed through conversational queries, supporting context-aware reconstruction of prior events.

No formal user study or empirical validation has been conducted at this stage; the present work establishes a functional proof-of-concept. The focus is on demonstrating the technical feasibility of the end-to-end capture-to-retrieval pipeline, including multimodal sensing, temporal synchronization, structured archival, and natural language query-based access. These results position CPMS as an operational prototype for AI-enabled episodic memory augmentation and provide a foundation for future user-centered evaluation \cite{akbari2021, gupta2022}.

\begin{figure}[H]
    \centering
    \includegraphics[width=\linewidth,keepaspectratio]{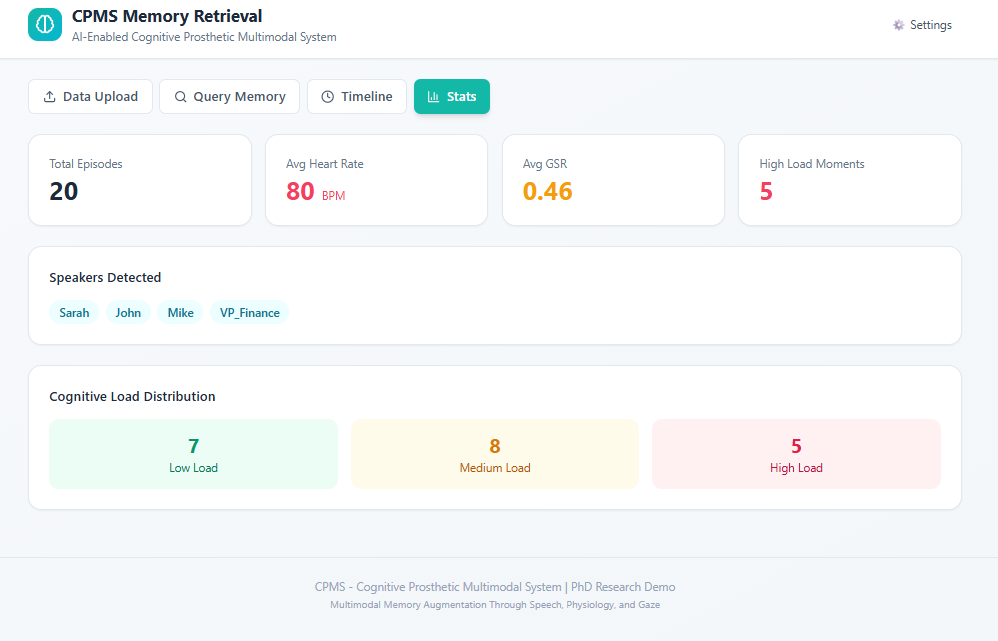}
    \Description{Screenshot of aggregated statistics panels summarizing episodic data.}
    \caption{Aggregate statistics derived from multimodal episodes.}
    \label{fig:cpms-stats}
\end{figure}

\begin{figure}[H]
    \centering
\includegraphics[width=\linewidth,keepaspectratio]{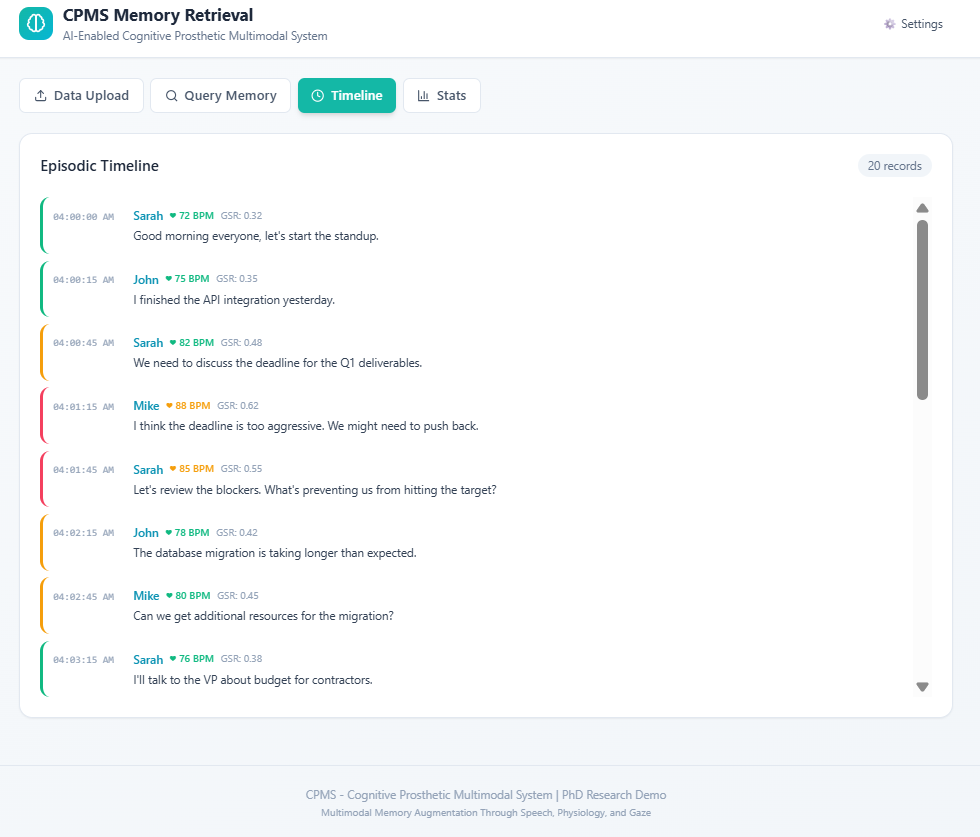}
    \Description{Screenshot of a timeline view showing episodic entries laid out over time.}
    \caption{Timeline-based navigation of episodic memory records.}
    \label{fig:cpms-timeline}
\end{figure}

The three figures illustrate key components of the CPMS memory retrieval interface. Figure~\ref{fig:cpms-interface} presents the natural language query interface, where users submit free-text questions to retrieve episodic records. Figure~\ref{fig:cpms-timeline} shows the timeline-based navigation view, enabling users to browse synchronized episodic entries across time. Figure~\ref{fig:cpms-stats} displays aggregated statistics derived from multimodal episodes, providing summarized insights into physiological, speech, and interaction patterns. Together, these views demonstrate how CPMS supports interactive, context-aware episodic recall through query-driven retrieval, temporal exploration, and summary-level analysis.

\section{Discussion}

The development of CPMS illustrates that multimodal sensing, structured archival storage, and AI-enabled retrieval can be feasibly integrated into a cohesive system targeting episodic memory support in workplace settings. Rather than contributing experimental performance metrics, this work shows that speech, physiological, and gaze data can be reliably captured, synchronized, serialized into coherent JSON-based episodic records, and accessed through a functional natural language retrieval interface. Together, these components establish CPMS as an end-to-end proof-of-concept for episodic memory augmentation, moving beyond data preparation to operational AI-mediated recall and reflection.

In contrast to single-modal memory aids or exhaustive lifelogging systems that primarily store raw audio or video streams \cite{epstein2020, cho2022}, CPMS produces a lightweight, multimodal representation that is both machine-readable and semantically interpretable. By transforming heterogeneous signals into temporally aligned, feature-level entries and exposing them through a queryable web application, CPMS advances toward cognitive scaffolds \cite{heersmink2022} that reflect human episodic segmentation while remaining privacy-aware and deployable. Positioned between unstructured lifelogging, semantic-only retrieval frameworks \cite{ribeiro2023}, and large-scale capture initiatives such as Ego4D \cite{grauman2022}, CPMS illustrates that selective, synchronized, and ethically grounded logging can support meaningful memory reconstruction without the burdens of continuous surveillance or high-bandwidth storage.

The structured multimodal logs and retrieval interface enable retrospective recall, contextual review, and attention-aware reflection, supporting natural language queries about what was said, attended to, or physiologically experienced during specific moments \cite{yang2021, zacks2020, izumi2021}. Episodic structuring aligns with psychological evidence that recall is driven by associative cues across modalities \cite{zacks2020, baldwin2020}, positioning CPMS as a transparent and auditable cognitive prosthetic rather than a black-box summarization tool. At its current stage, CPMS remains a technical proof-of-concept, with future work focusing on longitudinal deployments and user studies to assess recall utility, cognitive load, and trust in AI-mediated memory systems.

\subsection{Ethical Considerations and Privacy Safeguards}

While CPMS employs privacy-by-design through local preprocessing and on-device storage \cite{liu2023, badidi2023}, continuous workplace sensing introduces ethical considerations that extend beyond data locality. Workplace deployment of multimodal sensing raises concerns around informed consent, bystander capture, surveillance perception, and potential data misuse that must be addressed for responsible adoption.

To mitigate these risks, CPMS incorporates several design safeguards. The system operates on an explicit opt-in basis: sensing is activated only by the user and can be paused or terminated at any time through the interface, ensuring individual agency over data capture. All episodic records are stored locally and owned exclusively by the user, with no data transmitted to external servers or shared with employers without explicit consent. The modular architecture further supports selective modality capture, allowing users to disable specific sensors (e.g., physiological monitoring) based on comfort or organizational policy.

Important challenges remain, however. Bystander speech captured during meetings raises consent questions for non-users present in shared workspaces. Future deployments should implement speaker diarization to identify and optionally exclude bystander contributions \cite{song2021}, alongside clear signaling mechanisms (e.g., visible indicators that recording is active) to support informed consent among co-located individuals. Access control policies, data encryption at rest, and configurable retention and deletion schedules are planned for production deployments to support compliance with organizational and regulatory requirements. These safeguards position CPMS not as a surveillance tool, but as a user-controlled cognitive aid that prioritizes individual autonomy and transparency.

\section{Conclusion}

This paper introduced the CPMS, an AI-enabled proof-of-concept designed to support episodic memory, awareness, and reflection in knowledge-intensive workplace environments. CPMS demonstrates the technical feasibility of synchronizing heterogeneous signals—speech, physiological responses, and gaze behavior, into queryable JSON-based episodic records accessible through a functional natural language memory retrieval web application.

Rather than emphasizing exhaustive lifelogging or real-time intervention, CPMS prioritizes multimodal integration, contextual structuring, and interpretability. By abstracting raw sensor streams into feature-level representations such as aligned transcripts, stress indicators \cite{farrow2019technological}, and gaze fixations, the system enables privacy-conscious, semantically grounded memory reconstruction. Its modular design supports operation with partial sensor configurations, and its ethical safeguards, including opt-in activation, local-only storage, and user-owned data, position CPMS as a transparent cognitive scaffold that supports recall and reflection without imposing continuous surveillance, aligning with ethical requirements for workplace deployment.

The central contribution of this work lies in presenting an end-to-end data-to-memory system that bridges multimodal sensing, structured archival storage, and AI-enabled natural language retrieval. Through structured JSON encoding and an operational web-based query interface, CPMS advances a practical and ethically grounded approach to cognitive augmentation, illustrating how episodic memory support can be integrated into real-world professional settings. Formal user evaluation, including assessment of recall utility, cognitive load, and trust, remains future work essential to validating the system's impact on knowledge workers.

\section{Future Work}

While CPMS demonstrates the technical feasibility of multimodal episodic memory capture and AI-enabled retrieval, systematic user evaluation remains future work. Initial evaluation will follow a staged methodology: first, a small-scale pilot study (N=8--12) with knowledge workers performing simulated meeting and task-resumption scenarios will assess system usability, retrieval accuracy, and perceived recall benefit using standardized measures (e.g., System Usability Scale, NASA-TLX for cognitive load). 

The Findings of the pilot will inform iterative refinement of the retrieval interface and episodic structure before proceeding to a controlled within-subject study comparing the performance of recall and the efficiency of task-resumption with and without CPMS support. Subsequent studies will examine how different user populations interact with and benefit from episodic memory scaffolding, including individuals with attention-related challenges (e.g., ADHD), visually impaired users, autistic individuals, and neurotypical knowledge workers. These studies will assess recall utility, cognitive load, accessibility, and trust in AI-mediated memory systems across diverse cognitive and perceptual profiles.

Future extensions will also explore adaptive and accessible interface designs, interactive visualization dashboards for episodic exploration, and longitudinal deployments in individual and collaborative workplace settings. Additional work will investigate system robustness under degraded conditions, including noisy ASR input, absent sensor modalities, and physiological signal artifacts, to characterize retrieval reliability across realistic deployment scenarios. Further development will integrate new contextual sensors and leverage multimodal language models for richer episodic summarization and explanation. Together, these directions aim to advance CPMS from a proof-of-concept into a deployable, inclusive, and trustworthy cognitive prosthetic that augments, rather than replaces, human memory and sensemaking.


\bibliographystyle{ACM-Reference-Format}

\bibliography{references}

@inproceedings{cho2022,
author = {Cho, Janghee and Xu, Tian and Zimmermann-Niefield, Abigail and Voida, Stephen},
year = {2022},
month = {04},
pages = {1-23},
title = {Reflection in Theory and Reflection in Practice: An Exploration of the Gaps in Reflection Support among Personal Informatics Apps},
doi = {10.1145/3491102.3501991}
}

@inproceedings{thennakoon2025devices,
  title={How Devices Shape Mental Effort in Digital Document Reading: An Eye-Tracking Study},
  author={Thennakoon, Kumushini and Abeysinghe, Yasasi and Thenahandi, Pasindu and Obiuwevwi, Lawrence and Ashok, Vikas and Jayarathna, Sampath},
  booktitle={2025 ACM/IEEE Joint Conference on Digital Libraries (JCDL)},
  pages={237--240},
  year={2025},
  organization={IEEE}
}

@inproceedings{jayawardena2025real,
  title={A real-time approach to capture ambient and focal attention in visual search},
  author={Jayawardena, Gavindya and Jayawardana, Yasith and Abeysinghe, Yasasi and Mahanama, Bhanuka and Jayarathna, Sampath and Gwizdka, Jacek},
  booktitle={Proceedings of the 2025 Symposium on Eye Tracking Research and Applications},
  pages={1--7},
  year={2025}
}

@misc{obiuwevwi2025toward,
      title={Towards Affordable, Non-Invasive Real-Time Hypoglycemia Detection Using Wearable Sensor Signals}, 
      author={Lawrence Obiuwevwi and Krzysztof J. Rechowicz and Vikas Ashok and Sampath Jayarathna},
      year={2026},
      eprint={2602.10407},
      archivePrefix={arXiv},
      primaryClass={cs.HC},
      url={https://arxiv.org/abs/2602.10407}, 
}

@inproceedings{jayawardana2021metadata,
  title={Metadata-driven eye tracking for real-time applications},
  author={Jayawardana, Yasith and Jayawardena, Gavindya and Duchowski, Andrew T and Jayarathna, Sampath},
  booktitle={Proceedings of the 21st ACM Symposium on Document Engineering},
  pages={1--4},
  year={2021}
}

@inproceedings{jayawardana2019dfs,
  title={DFS: a dataset file system for data discovering users},
  author={Jayawardana, Yasith and Jayarathna, Sampath},
  booktitle={2019 ACM/IEEE Joint Conference on Digital Libraries (JCDL)},
  pages={355--356},
  year={2019},
  organization={IEEE}
}

@inproceedings{abeysinghe2025framework,
  title={Framework for measuring visual attention in gaze-driven vr learning environments using meta quest pro},
  author={Abeysinghe, Yasasi and Cauchi, Kevin and Ashok, Vikas and Jayarathna, Sampath},
  booktitle={Proceedings of the 2025 Symposium on Eye Tracking Research and Applications},
  pages={1--3},
  year={2025}
}

@inproceedings{jayawardana2022streaminghub,
  title={StreamingHub: interactive stream analysis workflows},
  author={Jayawardana, Yasith and Ashok, Vikas G and Jayarathna, Sampath},
  booktitle={Proceedings of the 22nd ACM/IEEE Joint Conference on Digital Libraries},
  pages={1--10},
  year={2022}
}

@inproceedings{farrow2019technological,
  title={Technological advancements in post-traumatic stress disorder detection: a survey},
  author={Farrow, Bathsheba and Jayarathna, Sampath},
  booktitle={2019 IEEE 20th International Conference on Information Reuse and Integration for Data Science (IRI)},
  pages={223--228},
  year={2019},
  organization={IEEE}
}

@inproceedings{jayarathna2017analysis,
  title={Analysis and modeling of unified user interest},
  author={Jayarathna, Sampath and Shipman, Frank},
  booktitle={2017 IEEE International Conference on Information Reuse and Integration (IRI)},
  pages={298--307},
  year={2017},
  organization={IEEE}
}

@inproceedings{jayarathna2016rationale,
  title={Rationale and architecture for incorporating human oculomotor plant features in user interest modeling},
  author={Jayarathna, Sampath and Shipman, Frank},
  booktitle={Proceedings of the 2016 ACM on Conference on Human Information Interaction and Retrieval},
  pages={281--284},
  year={2016}
}

@inproceedings{thennakoon2025beyond,
  title={Beyond gaze overlap: Analyzing joint visual attention dynamics using egocentric data},
  author={Thennakoon, Kumushini and Abeysinghe, Yasasi and Mahanama, Bhanuka and Ashok, Vikas and Jayarathna, Sampath},
  booktitle={2025 IEEE International Conference on Information Reuse and Integration and Data Science (IRI)},
  pages={337--342},
  year={2025},
  organization={IEEE}
}

@inproceedings{jayawardena2020pilot,
  title={Pilot study of audiovisual speech-in-noise (sin) performance of young adults with adhd},
  author={Jayawardena, Gavindya and Michalek, Anne and Duchowski, Andrew and Jayarathna, Sampath},
  booktitle={ACM symposium on eye tracking research and applications},
  pages={1--5},
  year={2020}
}

@article{de2019survey,
  title={A survey of attention deficit hyperactivity disorder identification using psychophysiological data},
  author={De Silva, Senuri and Dayarathna, Sanuwani and Ariyarathne, Gangani and Meedeniya, Dulani and Jayarathna, Sampath},
  year={2019},
  publisher={International Association of Online Engineering}
}

@article{zacks2020,
author = {Zacks, Jeffrey},
year = {2020},
month = {01},
pages = {165-191},
title = {Event Perception and Memory},
volume = {71},
journal = {Annual Review of Psychology},
doi = {10.1146/annurev-psych-010419-051101}
}

@article{baldwin2020,
  author    = {Baldwin DA, Kosie JE.},
  title     = {How Does the Mind Render Streaming Experience as Events?},
  journal   = {Topics in Cognitive Science},
  year      = {2020},
  doi       = {10.1111/tops.12502}
}

@article{heersmink2022,
  title={Preserving Narrative Identity for Dementia Patients: Embodiment, Active Environments, and Distributed Memory},
  author={Richard Heersmink},
  journal={Neuroethics},
  year={2022},
  volume={15},
  url={https://api.semanticscholar.org/CorpusID:246654498}
}

@inproceedings{yen2021,
  author    ={An-Zi Yen and Hen-Hsen Huang and Hsin-Hsi Chen},
  title     = {Ten Questions in Lifelog Mining and Information Recall},
  booktitle = {Proceedings of the 2021 International Conference on Multimedia Retrieval},
  year      = {2021},
  publisher = {ACM},
  doi       = {10.1145/3460426.3463607}
}

@inproceedings{song2021,
author = {Song, Yuanfeng and Jiang, Di and Zhao, Xuefang and Huang, Xiaoling and Xu, Qian and Wong, Raymond Chi-Wing and Yang, Qiang},
title = {SmartMeeting: Automatic Meeting Transcription and Summarization for In-Person Conversations},
year = {2021},
isbn = {9781450386517},
publisher = {Association for Computing Machinery},
address = {New York, NY, USA},
url = {https://doi.org/10.1145/3474085.3478556},
doi = {10.1145/3474085.3478556},
booktitle = {Proceedings of the 29th ACM International Conference on Multimedia},
pages = {2777–2779},
numpages = {3},
location = {Virtual Event, China},
series = {MM '21}
}

@ARTICLE{izumi2021, 
AUTHOR={Izumi, Keisuke  and Minato, Kazumichi  and Shiga, Kiko  and Sugio, Tatsuki  and Hanashiro, Sayaka  and Cortright, Kelley  and Kudo, Shun  and Fujita, Takanori  and Sado, Mitsuhiro  and Maeno, Takashi  and Takebayashi, Toru  and Mimura, Masaru  and Kishimoto, Taishiro },        
TITLE={Unobtrusive Sensing Technology for Quantifying Stress and Well-Being Using Pulse, Speech, Body Motion, and Electrodermal Data in a Workplace Setting: Study Concept and Design},   
JOURNAL={Frontiers in Psychiatry},
VOLUME={Volume 12 - 2021}, 
YEAR={2021},
URL={https://www.frontiersin.org/journals/psychiatry/articles/10.3389/fpsyt.2021.611243},
DOI={10.3389/fpsyt.2021.611243},
ISSN={1664-0640}}

@inproceedings{agustianto2022,
  title={Eye Tracking Usability Testing Using User-Centered Design Analysis Method},
  author={Khafidurrohman Agustianto and Adi Heru Utomo and Ratih Ayuninghemi and Prawidya Destarianto and I Gede Wiryawan and Ely Mulyadi},
  year={2022},
  booktitle={Proceedings of the 2nd International Conference on Social Science, Humanity and Public Health (ICOSHIP 2021)},
  pages={265-269},
  issn={2352-5398},
  isbn={978-94-6239-539-8},
  url={https://doi.org/10.2991/assehr.k.220207.045},
  doi={10.2991/assehr.k.220207.045},
  publisher={Atlantis Press}
}

@article{turkoglu2024,
author = {T\"{u}rko\u{g}lu, Hacer and Yal\c{c}\i{}nalp, Serpil},
title = {Investigating problem-solving behaviours of university students through an eye-tracking system using GeoGebra in geometry: A case study},
year = {2024},
issue_date = {Aug 2024},
publisher = {Kluwer Academic Publishers},
address = {USA},
volume = {29},
number = {12},
issn = {1360-2357},
url = {https://doi.org/10.1007/s10639-024-12452-1},
doi = {10.1007/s10639-024-12452-1},
journal = {Education and Information Technologies},
month = feb,
pages = {15761–15791},
numpages = {31},
}

@article{pourhosein2025,
author = {Pourhosein, Mahshid and Sabokro, Mehdi},
year = {2025},
month = {02},
pages = {},
title = {Unveiling the gaze: deciphering key factors in selecting knowledge workers through eye-tracking analysis},
volume = {30},
journal = {European Journal of Management Studies},
doi = {10.1108/EJMS-10-2024-0106}
}

@article{tilkar2025a,
author = {Tilkar, Swati},
year = {2025},
month = {06},
pages = {1-9},
title = {Generating Meeting Transcription Using Natural Language Processing},
volume = {09},
journal = {International Journal of Scientific Research in Engineering and Management},
doi = {10.55041/IJSREM51091}
}

@article{midha2020,
author = {Midha, Serena and Maior, Horia and Wilson, Max and Sharples, Sarah},
year = {2021},
month = {03},
pages = {102580},
title = {Measuring Mental Workload Variations in Office Work Tasks using fNIRS},
volume = {147},
journal = {International Journal of Human-Computer Studies},
doi = {10.1016/j.ijhcs.2020.102580}
}

@article{wang2021,
author = {Wang, Jiyang and Grant, Trevor and Gursoy, Senem and Geng, Baocheng and Hirshfield, Leanne},
year = {2021},
month = {12},
pages = {1-1},
title = {Taking a Deeper Look at the Brain: Predicting Visual Perceptual and Working Memory Load from High-Density fNIRS Data},
volume = {PP},
journal = {IEEE Journal of Biomedical and Health Informatics},
doi = {10.1109/JBHI.2021.3133871}
}

@article{colabianchi2024,
author = {Colabianchi, Silvia and Costantino, Francesco and Sabetta, Nicol\`{o}},
title = {Assessment of a large language model based digital intelligent assistant in assembly manufacturing},
year = {2024},
issue_date = {Nov 2024},
publisher = {Elsevier Science Publishers B. V.},
address = {NLD},
volume = {162},
number = {C},
issn = {0166-3615},
url = {https://doi.org/10.1016/j.compind.2024.104129},
doi = {10.1016/j.compind.2024.104129},
journal = {Comput. Ind.},
month = nov,
numpages = {21}
}

@article{duchevet2022,
author = {Duchevet, Alexandre and Imbert, Jean-Paul and de la Hogue, Théo and Ferreira, A. and Moens, Laura and Colomer, Adrián and Cantero, J. and Bejarano, C. and Vázquez, A.},
year = {2022},
month = {12},
pages = {253-261},
title = {HARVIS: a digital assistant based on cognitive computing for non-stabilized approaches in Single Pilot Operations},
volume = {66},
journal = {Transportation Research Procedia},
doi = {10.1016/j.trpro.2022.12.025}
}

@article{brynjolfsson2023,
author = {Brynjolfsson, Erik and Li, Danielle and Raymond, Lindsey},
year = {2023},
month = {04},
journal   = {NBER Working Paper},
pages = {},
title = {Generative AI at Work},
doi = {10.48550/arXiv.2304.11771}
}

@article{ding2022,
  title={Device-Free Multi-Location Human Activity Recognition Using Deep Complex Network},
  author={Xue Ding and Chunlei Hu and Weiliang Xie and Yi Zhong and Jianfei Yang and Ting Jiang},
  journal={Sensors (Basel, Switzerland)},
  year={2022},
  volume={22},
  url={https://api.semanticscholar.org/CorpusID:251703766}
}

@inproceedings{ribeiro2023,
author = {Ribiero, Ricardo and Trifan, Alina and Neves, Antonio J. R.},
title = {MEMORIA: A Memory Enhancement and MOment RetrIeval Application for LSC 2022},
year = {2022},
isbn = {9781450392396},
publisher = {Association for Computing Machinery},
address = {New York, NY, USA},
url = {https://doi.org/10.1145/3512729.3533011},
doi = {10.1145/3512729.3533011},
booktitle = {Proceedings of the 5th Annual on Lifelog Search Challenge},
pages = {8–13},
numpages = {6},
location = {Newark, NJ, USA},
series = {LSC '22}
}

@inproceedings{Hodges2006,
author = {Hodges, Steve and Williams, Lyndsay and Berry, Emma and Izadi, Shahram and Srinivasan, James and Butler, Alex and Smyth, Gavin and Kapur, Narinder and Wood, Ken},
title = {SenseCam: a retrospective memory aid},
year = {2006},
isbn = {9783540396345},
publisher = {Springer-Verlag},
address = {Berlin, Heidelberg},
url = {https://doi.org/10.1007/11853565_11},
doi = {10.1007/11853565_11},
booktitle = {Proceedings of the 8th International Conference on Ubiquitous Computing},
pages = {177–193},
numpages = {17},
location = {Orange County, CA},
series = {UbiComp'06}
}

@INPROCEEDINGS{obiuwevwi2025hypoglycemia,
  author       = {Obiuwevwi, Lawrence and Rechowicz, Krzysztof J. and Ashok, Vikas and Jayarathna, Sampath},
  title        = {Toward Affordable and Non-Invasive Detection of Hypoglycemia: A Machine Learning Approach},
  booktitle    = {Proceedings of the 2025 IEEE International Conference on Information Reuse and Integration and Data Science (IRI)},
  year         = {2025},
  pages        = {156--161},
  doi          = {10.1109/IRI66576.2025.00036},

}

@ARTICLE{belkacem2020,
  author    = {Belkacem, Abdelkader Nasreddine},
  title     = {Brain Computer Interfaces for Improving the Quality of Life of Older Adults and Elderly Patients},
  journal   = {Frontiers in Neuroscience},
  year      = {2020},
  volume    = {14},
  pages     = {692},
  doi       = {10.3389/fnins.2020.00692}
}

@INPROCEEDINGS{grauman2022,
  author={Grauman, Kristen and Westbury, Andrew and Byrne, Eugene and Chavis, Zachary and Furnari, Antonino and Girdhar, Rohit and Hamburger, Jackson and Jiang, Hao and Liu, Miao and Liu, Xingyu and Martin, Miguel and Nagarajan, Tushar and Radosavovic, et al},
  booktitle={2022 IEEE/CVF Conference on Computer Vision and Pattern Recognition (CVPR)}, 
  title={Ego4D: Around the World in 3,000 Hours of Egocentric Video}, 
  year={2022},
  volume={},
  number={},
  pages={18973-18990},
  doi={10.1109/CVPR52688.2022.01842}}

@ARTICLE{zhou2024,
  author    = {Yuchen Zhou and Wenqi Shao and Dahua Lin},
  title     = {Learning from Observer Gaze: Zero-Shot Attention Prediction Oriented by Human-Object Interaction Recognition},
  journal   = {arXiv preprint},
  year      = {2024},
  volume    = {abs/2405.09931},
  doi       = {10.48550/arxiv.2405.09931}
}

@INPROCEEDINGS{Kumushini2025,
  author={Thennakoon, Kumushini and Abeysinghe, Yasasi and Thenahandi, Pasindu and Obiuwevwi, Lawrence and Ashok, Vikas and Jayarathna, Sampath},
  booktitle={2025 ACM/IEEE Joint Conference on Digital Libraries (JCDL)}, 
  title={How Devices Shape Mental Effort in Digital Document Reading: An Eye-Tracking Study}, 
  year={2025},
  pages={237-240},
  doi={10.1109/JCDL67857.2025.00035}}

@inproceedings{akbari2021,
author = {Akbari, Hassan and Yuan, Liangzhe and Qian, Rui and Chuang, Wei-Hong and Chang, Shih-Fu and Cui, Yin and Gong, Boqing},
title = {VATT: transformers for multimodal self-supervised learning from raw video, audio and text},
year = {2021},
isbn = {9781713845393},
publisher = {Curran Associates Inc.},
address = {Red Hook, NY, USA},
booktitle = {Proceedings of the 35th International Conference on Neural Information Processing Systems},
articleno = {1853},
numpages = {16},
series = {NIPS '21}
}

@misc{xu2023multimodal,
      title={Multimodal Learning with Transformers: A Survey}, 
      author={Peng Xu and Xiatian Zhu and David A. Clifton},
      year={2023},
      eprint={2206.06488},
      archivePrefix={arXiv},
      primaryClass={cs.CV},
      url={https://arxiv.org/abs/2206.06488}, 
}

@article{dobbins2017detecting,
  author    = {Chelsea Dobbins and Reza Rawassizadeh and Elaheh Momeni},
  title     = {Detecting Physical Activity Within Lifelogs Towards Preventing Obesity and Aiding Ambient Assisted Living},
  journal   = {Neurocomputing},
  year      = {2017},
  volume    = {230},
  pages     = {110--132},
  issn      = {0925-2312},
  doi       = {10.1016/j.neucom.2016.02.088},
}

@Article{schmidt2021,
AUTHOR = {Dritsas, Elias and Trigka, Maria and Troussas, Christos and Mylonas, Phivos},
TITLE = {Multimodal Interaction, Interfaces, and Communication: A Survey},
JOURNAL = {Multimodal Technologies and Interaction},
VOLUME = {9},
YEAR = {2025},
NUMBER = {1},
ARTICLE-NUMBER = {6},
URL = {https://www.mdpi.com/2414-4088/9/1/6},
ISSN = {2414-4088},
DOI = {10.3390/mti9010006}
}

@Article{gupta2022,
AUTHOR = {Lahiri, Aritra Kumar and Hu, Qinmin Vivian},
TITLE = {AlzheimerRAG: Multimodal Retrieval-Augmented Generation for Clinical Use Cases},
JOURNAL = {Machine Learning and Knowledge Extraction},
VOLUME = {7},
YEAR = {2025},
NUMBER = {3},
ARTICLE-NUMBER = {89},
URL = {https://www.mdpi.com/2504-4990/7/3/89},
ISSN = {2504-4990}}

@inproceedings{Marusich2025,
  title={Trust Calibration for Joint Human/AI Decision-Making in Dynamic and Uncertain Contexts},
  author={Laura Marusich and Benjamin T. Files and Melanie Bancilhon and Justine Rawal and Adrienne Raglin},
  booktitle={Interacci{\'o}n},
  year={2025},
  url={https://api.semanticscholar.org/CorpusID:279265662}
}

@ARTICLE{liu2023,
  author={Liu, X. and Chakraborty, B.},
  journal={American Journal of Public Health},
  title={Microrandomized Trials: Developing Just-in-Time Adaptive Interventions},
  year={2023},
  volume={113},
  number={1},
  pages={60–69},
  doi={10.2105/AJPH.2022.307150}
}

@article{ali2022,
author = {Ali, Shaukat and Khusro, Shah and Khan, Akif and Khan, Inayat and Alam, Iftikhar and Solehria, Salman},
year = {2022},
month = {12},
pages = {1-32},
title = {SLOG: smartphone-based semantic lifelogging framework for digital prosthetic memory development},
volume = {27},
journal = {Personal and Ubiquitous Computing},
doi = {10.1007/s00779-022-01701-0}
}

@ARTICLE{gonzalez2024,
  author={Gonz{\'a}lez-D{\'i}az, I. and Molina-Moreno, M. and Benois-Pineau, J. and de Rugy, A.},
  journal={IEEE Journal of Biomedical and Health Informatics}, 
  title={Asymmetric Multi-Task Learning for Interpretable Gaze-Driven Grasping Action Forecasting}, 
  year={2024},
  volume={28},
  number={12},
  pages={7517--7530},
  month={Dec},
  doi={10.1109/JBHI.2024.3430810}
}

@article{laing2025,
author = {Laing, Patrick and Dunsmoor, Joseph},
year = {2025},
month = {01},
pages = {110-134},
title = {Event Segmentation Promotes the Reorganization of Emotional Memory},
volume = {37},
journal = {Journal of Cognitive Neuroscience},
doi = {10.1162/jocn_a_02244}
}

@article{li2025,
author = {Li, Yue and Johansson, Mikael and Nikolaev, Andrey},
year = {2025},
month = {05},
pages = {},
title = {Hierarchical event segmentation of episodic memory in virtual reality},
volume = {10},
journal = {npj Science of Learning},
doi = {10.1038/s41539-025-00321-6}
}

@article{calvo2021affect,
author = {Calvo, Rafael and Peters, Dorian},
year = {2016},
month = {12},
pages = {84-85},
title = {Positive Computing: Technology for Wellbeing and Human Potential},
volume = {22},
journal = {Psychology Teaching Review},
doi = {10.53841/bpsptr.2016.22.2.84}
}

@ARTICLE{badidi2023,
  author={Badidi, Elarbi and Moumane, Karima and Ghazi, Firdaous El},
  journal={IEEE Access}, 
  title={Opportunities, Applications, and Challenges of Edge-AI Enabled Video Analytics in Smart Cities: A Systematic Review}, 
  year={2023},
  volume={11},
  pages={80543-80572},
  doi={10.1109/ACCESS.2023.3300658}}

@article{epstein2020,
author = {Epstein, Daniel A. and Caldeira, Clara and Figueiredo, Mayara Costa and Lu, Xi and Silva, Lucas M. and Williams, Lucretia and Lee, Jong Ho and Li, Qingyang and Ahuja, Simran and Chen, Qiuer and Dowlatyari, Payam and Hilby, Craig and Sultana, Sazeda and Eikey, Elizabeth V. and Chen, Yunan},
title = {Mapping and Taking Stock of the Personal Informatics Literature},
year = {2020},
issue_date = {December 2020},
publisher = {Association for Computing Machinery},
address = {New York, NY, USA},
volume = {4},
number = {4},
url = {https://doi.org/10.1145/3432231},
doi = {10.1145/3432231},
journal = {Proc. ACM Interact. Mob. Wearable Ubiquitous Technol.},
month = dec,
articleno = {126},
numpages = {38}
}

@article{yang2021,
author = {Yang, Po and Bi, Gaoshan and Qi, Jun and Wang, Xulong and Yang, Yun and Xu, Lida},
title = {Multimodal Wearable Intelligence for Dementia Care in Healthcare 4.0: a Survey},
year = {2021},
issue_date = {Feb 2025},
publisher = {Kluwer Academic Publishers},
address = {USA},
volume = {27},
number = {1},
issn = {1387-3326},
url = {https://doi.org/10.1007/s10796-021-10163-3},
doi = {10.1007/s10796-021-10163-3},
journal = {Information Systems Frontiers},
month = jul,
pages = {197–214},
numpages = {18}
}

\end{document}